\documentclass[twocolumn,pre,showkeys,amsmath,amssymb,superscriptaddress,showpacs,floatfix]{revtex4}
\usepackage[dvips]{graphicx}  
\begin{document}

\title[Short-time dynamic in the Majority vote model: The ordered and disordered initial cases]
{Short-time dynamic in the Majority vote model: The ordered and disordered initial cases}

\author{Francisco Sastre}

\affiliation{Departamento de Ingenier\'ia F\'isica, Divisi\'on de Ciencias e Ingenier\'ia, Campus Le\'on de la Universidad de Guanajuato, AP E-143,
CP 37150, Le\'on, Guanajuato, M\'exico}
\email{sastre@fisica.ugto.mx}
\begin{abstract}
This work presents short-time Monte Carlo simulations for the
two dimensional Majority-vote model starting from ordered and disordered states.
It has been found that there are two pseudo-critical points, each one within the
error-bar range of previous reported values performed using fourth order
cumulant crossing method. The results show that the short-time dynamic for this model
has a dependence on the initial conditions. Based on this dependence a 
method is proposed for the evaluation of the pseudo critical points and the extraction of the dynamical 
critical exponent $z$ and the static critical exponent $\beta/\nu$ for this model.
\end{abstract}

\maketitle

\section{Introduction}
Critical phenomena in equilibrium statistical systems is one of the most
important topics in physics. Much of the attention has been focused on
the universality of the critical exponents, with several universality classes
already characterized in equilibrium systems. On the other hand the critical behavior 
of non-equilibrium statistical systems has been receiving a lot of attention 
in recent years, but the characterization of the different universality classes 
is far from be complete. 
One of the simplest non-equilibrium models is the two dimensional majority vote model, 
an Ising-like system (up-down symmetry and spin-flip dynamic) with a continuous order-disorder transition
with the same critical exponents that the two dimensional Ising model
\cite{Oliveira91,Oliveira93,Kwak2007}, as expected from the prediction of Grinstein {\em et al}
\cite{Grinstein85}: every spin system with spin-flip dynamic and up-down symmetry falls
in the Ising model universality class. However, there is some controversy about the universality class
for higher dimensions. A recent work claims~\cite{Yang2008} that the upper critical dimension for the
majority vote model is  6 instead of 4, based on numerical calculations.
Another discrepancy in the critical exponents have been found in simulations 
on non-regular lattices~\cite{Lima2005,Lima2006}.
It must be mentioned that all of the results mentioned above were performed
using standard "Monte Carlo" simulations and Finite Size Scaling approaches
for the evaluation of the static critical exponents.
On the other hand the time evolution can gives important
information about the universality of a given system. It has been shown by Janssen 
{\em et al.}~\cite{janssen89} that when systems with
relaxation dynamics are quenched from high temperatures to the
critical temperature there is a short critical universal behavior.
Numerical simulations have confirmed this behavior in the Ising and the Pott
models (see reference~\cite{zheng98} for a review of these results).
Concerning the critical dynamic in systems without detailed balance,
there are some works
that evaluate the critical dynamic exponent $z$~\cite{mendes98,tome98}
and the fluctuation-dissipation ratio $X_\infty$~\cite{sastre03} for the Majority
model, but always starting from a disordered state.
As expected the results were compatibles with the Ising ones. 
Given all these results
one should expect that the basic 
assumption for the dynamic relaxation of the k-th moment of the order parameter
starting from an completely ordered state will be the same as in the Ising
model, but this has not been proved yet.

The aims of this work are:
a) to evaluate the critical point using short time dynamic starting from
ordered and disordered initial conditions and b) evaluate the dynamic critical
exponent $z$ and the static critical exponent $\beta/\nu$.
This will test if the Grinstein prediction holds for the
short time dynamic in the majority vote model.

\section{Models and definitions}

In the Majority vote model each lattice site
has a spin whose values are $\sigma=\pm 1$ and its dynamic can be grouped by the spin flip rule
\begin{equation}
\label{dynamic}
W_i=\frac{1}{2}[1-\sigma_if(H_i)],
\end{equation}
here $H_i$ is the local field $\sum_\mathrm{nn}\sigma_j=,0,\pm 2,\pm 4$ produced
by its nearest neighbors and $f$ is a function with up down symmetry that 
depends on two control parameters $f(0)=0$, $f(2)=-f(-2)=x$ and $f(4)=-f(4)=y$. 
The parameters $x$ and $y$ can be associated with interface and bulk temperatures
respectively~\cite{drouffe99} using the relations
\begin{equation}
x=\tanh 2\beta_2,~~~~~~ y=\tanh 4\beta_4 .
\end{equation}
The Majority model is obtained setting $x=y$ ($\beta_4 < \beta_2$)and the critical point is at
$x_c=0.850(2)$~\cite{Oliveira93,Kwak2007}. The equilibrium case can be obtained
along the line $y=2x/(1+x^2)$ (Glauber dynamic), where the temperatures are equal ($\beta_2 =\beta_4 = \beta$)
and the critical point is $\beta_c = \frac{1}{2} \log(1+\sqrt{2})$.

The order parameter is the standard for Ising-like systems, defined by
\begin{equation}
m=\frac{1}{N}\langle\sum_i\sigma_i\rangle
\end{equation}
where $N=L^2$ is the total number of lattice sites ($L$ is the lateral size.

Starting from a disordered state the dynamic for the $k$-th moment of the
order parameter was derived by Janssen  {\em et al}~\cite{janssen89}, the mathematical expression is 
given by
\begin{equation}
m^{(k)}(t,\tau,L,m_0)=b^{-k\beta/\nu}m^{(k)}(b^{-z}t,b^{1/\nu}\tau,b^{-1}L,b^{x_0}m_0),
\end{equation}
here $\tau=(T-T_c)/T_c$ is the reduced temperature, or the reduced control parameter
for non-equilibrium systems, $t$ is the dynamic time
variable, $b$ is the re-scaling factor, $\beta$ and $\nu$ are static
critical exponents, $z$ is the dynamic critical exponent,$x_0$ is the scaling dimension
of the initial (small) order parameter $m_0$ (see~\cite{zheng98} and
reference therein).

At the critical point for sufficiently small $m_0$ and large systems ($L\to\infty$)  
the order parameter follows a power law dynamic
\begin{equation}
m(t)\sim m_0 t^{\theta},
\end{equation}
where $\theta$ is defined by $(x_0 - \beta/\nu)/z$. One must remark that there is a strong dependence
on the initial value of the order parameter. This dependence does not affect the power law behavior at the critical point,
what is affected is the exponent $\theta$ that tends to the real value in the limit $m_0\to 0$

For the dynamic starting, from the ordered state we have the assumption that the scaling dynamic
form is given by
\begin{equation}
m^{(k)}(t,\tau,L)=b^{-k\beta/\nu}m^{(k)}(b^{-z}t,b^{1/\nu}\tau,b^{-1}L),
\end{equation}
again it can be obtained the scaling form at the critical point taking the limit
$L\to \infty$.
\begin{equation}
M(t)\sim t^{-\beta/\nu z}.
\end{equation}

This has been proved in equilibrium systems, like the Ising or the Potts model.

For the evaluation of the critical point
it has been used the fact that theoretically the order parameter evolves as a 
power law at the critical point
If it is evaluated the difference between a power law and the time evolution for 
different values of the control parameter $x$ we will expect a minimum in those differences
at the critical point.
  
The simulations were performed 
choosing the lattice site randomly starting from both, a completely ordered state ($m=1.0$)
and a carefully prepared disordered state with small magnetization values $m_0 < 0.1$.
The order parameter is evaluated as a function of the time
$t$, with $\Delta t=1$ corresponding to a Monte Carlo time step (MCTS).
In order to avoid finite size effects we used lattices with lateral size
$L=2^{9}$ ($N=L^2$). The average were taken with at least $10^5$ independent simulations
and 250 and 1000 MCTS for the evaluation of the critical point and dynamical exponents
respectively.

\section{Results}

\begin{figure}
\begin{center}
\includegraphics[width=8cm,clip]{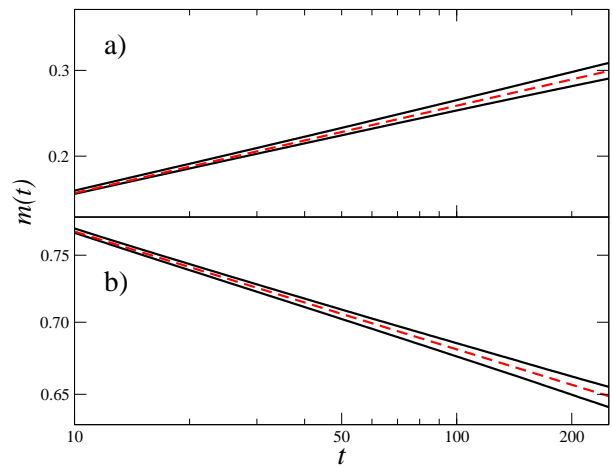}
\caption{\label{examples_dynamic} (color online) a) Dynamic starting from an ordered state.
b) Dynamic starting from a disordered state ($m_0 = 0.0875$). In both cases the top continuous 
line corresponds to a $x$ above the critical point and the lower one to a $x$ below the critical 
point. The dashed line shows the expected power law behavior.}
\end{center}
\end{figure}

Given that the power law behavior of the order parameter as function of $t$ (decay from an
ordered state, increase for a disordered one) one can performed simulations for different
values of $x$ around the expected critical point. In Fig.~\ref{examples_dynamic} we can observe
that above and below a certain value of $x$ we have dynamic that differ from a power law
(illustrated by the dashed line).

\begin{figure}
\begin{center}
\includegraphics[width=8cm,clip]{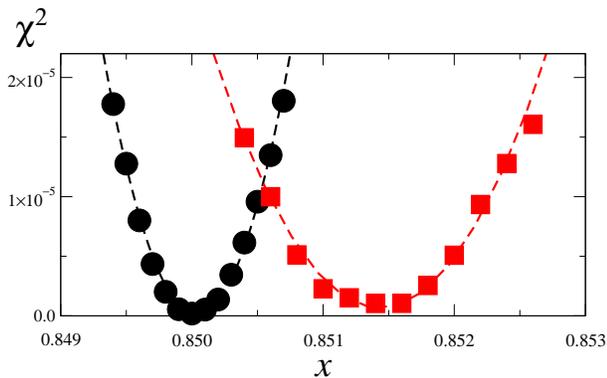}
\caption{\label{chisqr} (color online) Deviation from the power line behavior for the decay
($m_0=1$, left curve) and the increase ($m_0=0.0875$, right curve) cases.}
\end{center}
\end{figure}

From here one must define the criterion that will measure which is the best power law curve, in this work
it was used the measurement of the $\chi^2$ for each curve with respect to a power law behavior. I must remark that
previous results for equilibrium systems give the same result for the critical point for both
initial states. However, two different values were found for the majority vote model,
see Fig.~\ref{chisqr}.

The critical point values obtained were $x_c=0.85007(6)$ and $x_c=0.85147(2)$ for the order and disorder state
respectively, both results are in perfect agreement with the obtained in the static case (references~\cite{Oliveira91,Oliveira93,Kwak2007}) and are quite close.

We have a clearly dependence on the initial condition in short time dynamic of this model, which is not
the case for equilibrium systems. There is no doubt about the critical point obtained from the ordered phase,
since $m_0=1$ is one of the fixed point under renormalization group transformations. 
However, one must remember that the
power law is valid only in the limit $m_0 \to 0$ (the other fixed point) for the disorder case, 
assuming that the "real" critical point is located at this limit, one can proceed to evaluate 
the critical point for another values of $m_0$ and with this values extrapolate the critical point for the 
disordered phase. The results are showed in table\ref{critical_1}

\begin{center}
\begin{table}
\caption{\label{critical_1} Critical point for initial disordered states.}
\begin{tabular}{c|ccccc}
$m_0$ & 0.0375& 0.0500 & 0.0625 & 0.0750 & 0.0875 \\
\hline
$x_c$ & 0.84929(10) & 0.84972(9) & 0.85019(6) & 0.85076(5) & 0.85147(2) \\
\end{tabular}
\end{table}
\end{center}

\begin{figure}
\begin{center}
\includegraphics[width=8cm,clip]{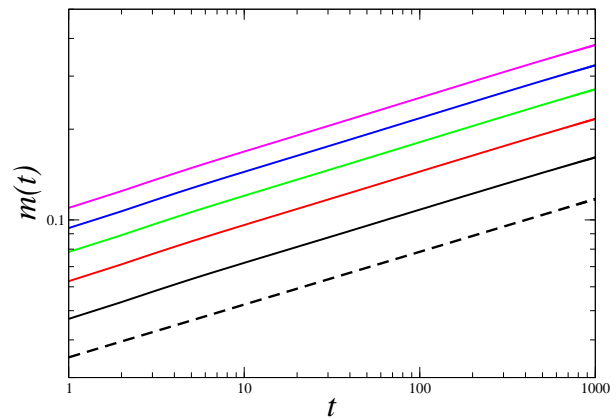}
\caption{\label{growing} (color online) Growing of the order parameter at $x_c(m_0)$, the continuous curves are
for $m_0=0.0875,~0.075,~0.0625,~0.5,~0.0375$ from top to bottom. The dashed line represents the
power law growing with $\theta=0.1751$ (the result in this work for the majority vote model).
}
\end{center}
\end{figure}

Once that each value $x_c(m_0)$ has been evaluated, the dynamical exponent $\theta$ can be obtained. 
For the evaluation of this exponent the simulations were performed with 1000 MCTS discarding
the first time steps, since there is an initial time scale $t_{mic}$ where the power law stabilizes $t_{mic}\sim 20$ (see Fig.~\ref{growing}). The results for the exponent $\theta$
are showed in table~\ref{critical_values}.

\begin{center}
\begin{table}
\caption{\label{critical_values}  $theta$ exponents for growing process.}
\begin{tabular}{c|ccccc}
$m_0$ & 0.0375& 0.0500 & 0.0625 & 0.0750 & 0.0875 \\
\hline
$\theta$ & 0.1769(8) & 0.1774(4) & 0.1782(3) & 0.1788(4) & 0.1792(3) \\
\end{tabular}
\end{table}
\end{center}

\begin{figure}
\centering
\includegraphics[width=8cm,clip]{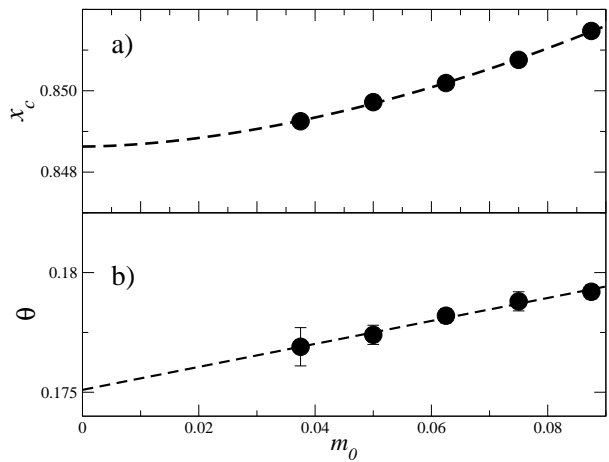}
\caption{\label{xc_theta} a) Evaluation of the critical point $x_c$ and b) the $\theta$ exponent.
}
\end{figure}

With a extrapolation of these values to $m_0=0$ the value of the critical point and the $\theta$ exponent were evaluated
(see Fig.~\ref{xc_theta}). The result for the critical point was $x_c=0.84860(10)$, which is clearly different from
the ordered one, however both values are within the error bar from the obtained in the static case $0.848\le x_c\le 0.852$.
From now on the pseudo critical point evaluated with the decay process will be denoted as $x_c^o$ and the evaluated
with the growing process with $x_c^d$. 
This surprising result seems similar to the obtained for weak first-order phase transitions~\cite{schulke},
where two pseudo-critical points exits due to the metastable states above and below the critical point.
However, there is an important difference in this case: for weak order phase transitions the smaller critical
point corresponds to the decay process, and the bigger corresponds to the growing process, contrary to the
majority vote model case. 

In the evaluation of the exponent $\theta$ a linear extrapolation gives the result of $0.175(3)$, which is 
lower from the values of the two dimensional Ising model, $\theta=0.191(1)$, and from the previously evaluated in references~\cite{mendes98,tome98} for the majority vote model, $\theta=0.192(2)$. The difference with respect to the Ising model could be understood considering
that the results obtained here seems to indicate a hole new dynamic. The differences
with previous results for the majority model can be explained observing the simulations details used previously:
first the critical point used was $x_=0.850$, which is above the result for $x_c^d$. Second the systems sizes used 
previously were really small ($L=32$), at this size the growing process is not very long and is really hard to
see the power law behavior.

\begin{figure}
\centering
\includegraphics[width=8cm,clip]{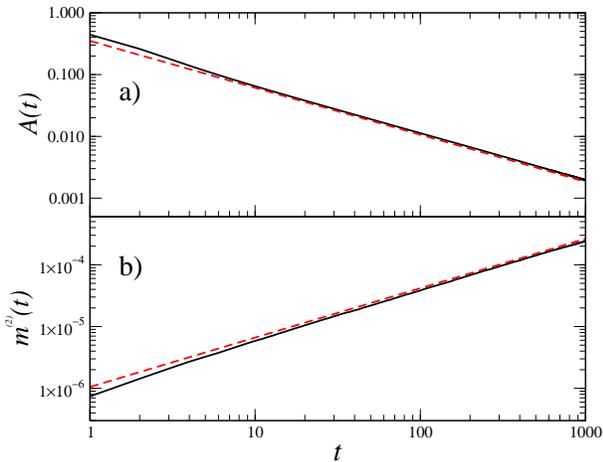}
\caption{\label{auto} (color online) a) Evaluation of $\lambda$, 
the continuous line shows the autocorrelation time evolution and the dashed line
shows the power law behavior with $\lambda=0.758$. b) Evaluation of $y$ ,
the continuous line shows the second moment order parameter and the dashed line
shows the power law behavior with $y=0.799$.
}
\end{figure}

One can obtain the dynamical exponent $z$ evaluating the second moment of the magnetization at the critical point $x_c^d$
\begin{equation}
m^{(2)}\sim t^y,~~~~~y=(d-2\beta/\nu)/z,
\end{equation}
and the autocorrelation
\begin{eqnarray}
\begin{array}{c}
A(t)=\sum_i \sigma_i(t=0)\sigma(t), \\
\\
A(t) \sim t^{-\lambda},~~~~~\lambda=\frac{d}{z} -\theta.
\end{array}
\end{eqnarray}
Both starting from $m_0=0$ and using 1000 MCTS. Again there is a $t_{mic}$ in each case (around 20 for the
autocorrelation and 75 for the second moment, see Fig.~\ref{auto}). The results obtained are $y=0.799(17)$ and
$\lambda=0.758(2)$. Combining both results it can be obtained the values $z=2.143(9)$ and $\beta/\nu = 0.143(18)$. A summary of the results are showed in table~\ref{summary}, where it can be observed discrepancies between most of the values for the majority model and the Ising ones. It must be remark that all these results were obtained using just the growing
process.

\begin{center}
\begin{table}
\caption{\label{summary} Summary of the results in this work and of the Ising model.}
\begin{tabular}{c|c|c}
~ & Majority vote model & Ising \\
\hline
$\theta$ & 0.175(3) & 0.191(1) \\
$\lambda$ & 0.758(2) & 0.737(1) \\
$z$ & 2.143(9) & 2.155(3) \\
$y$ & 0.799(17) &  0.817(7) \\
$\beta/\nu$ & 0.143(18) & 1/4 \\
\end{tabular}
\end{table}
\end{center}

\begin{figure}
\centering
\includegraphics[width=8cm,clip]{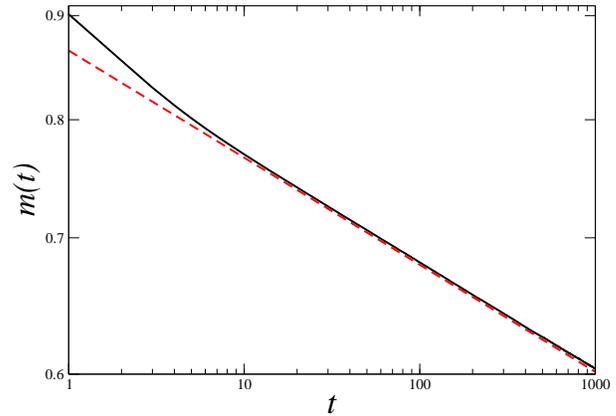}
\caption{\label{mag_decay} (color online) Order parameter relaxation ($m_0=1$), here the dashed line
shows the power law behavior with $\beta/\nu=0.0526$.
}
\end{figure}

Finally the decay exponent $\beta/\nu z$ was evaluated starting from an ordered phase (at $x_c^o$), using 1000 MCTS (Fig.~\ref{mag_decay}), the result was $\beta/\nu=0.0526(5)$, that is lower compared to the Ising one, $0.0580(5)$. Again the first time steps were discarded
for the evaluation of the exponent (Fig.~\ref{mag_decay}). Theoretically it is possible to obtain the $z$ exponent
using the known value of $\beta/\nu$, or knowing the $z$ value one can obtain the $\beta/\nu$ value, but in both cases
the results depend on values obtained with a growing process ($z$) or the static simulations ($\beta/nu$).  
The approach taken in this work is that the growing and the decay process are different and it could be possible that
the dynamic exponent $z$ is different in each case, so for the decay case the reporting value is $z=2.37(2)$.

\begin{figure}
\centering
\includegraphics[width=8cm,clip]{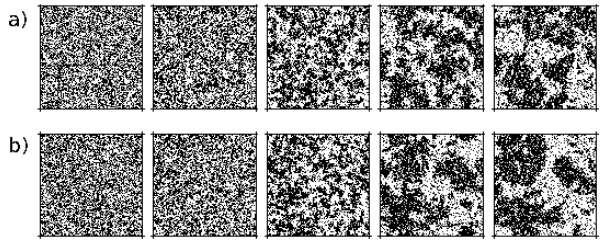}
\caption{\label{snapshot_dis} 
Snapshots at the pseudo-critical points starting with $m_0=0$, a) $x_c^d$ and b) $x_c^o$.
Times are (from left to right) 1, 100, 1000, 10000 and 20000.
}
\end{figure}

The fact that we have two pseudo-critical points (that not corresponds to a weak phase transition) and that
the decay and growing process are slower that in the Ising model must be related to the absence of detailed
balance condition. One of the consequences of this absence is that we do not have a unique thermodynamic
temperature, in this case we have two, so looking at the snapshots
for different initial conditions at the pseudo-critical points we can speculate about the competition between the
two "temperatures" that governs the dynamic in non-equilibrium Ising systems. 
Figure~\ref{snapshot_dis} shows the time evolution with $m_0=0$ at the two pseudo-critical points, a) for $x_c^d$ and
b) for $x_c^o$. Initially the number of sites with spin-flip probability depending on $\beta_2$ are very similar to the
ones depending on $\beta_4$, the time increases from left to right and it can be observed that at $x_c^o$ the coarsening
seems to appear faster that at $x_c^d$.

\begin{figure}
\centering
\includegraphics[width=8cm,clip]{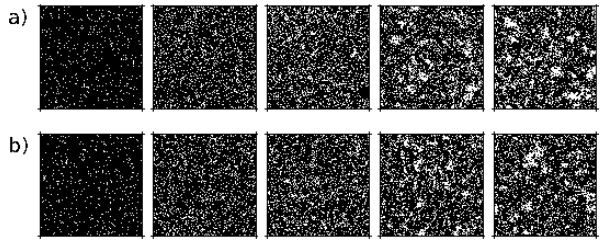}
\caption{\label{snapshot_ord}
Snapshots at the pseudo-critical points starting with $m_0=1$, a) $x_c^d$ and b) $x_c^o$.
Times are (from left to right) 1, 100, 1000, 10000 and 20000.
}
\end{figure}

Figure~\ref{snapshot_ord} shows the same time evolution for $m_0=1$. In this case at the beginning of the
evolution all sites have spin-flip probability that depends only on $\beta_4$ and the decay is
slightly lower at $x_c^o$.

\begin{figure}
\centering
\includegraphics[width=8cm,clip]{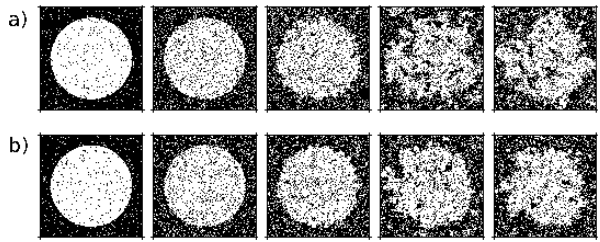}
\caption{\label{snapshot_cir} 
Snapshots at the pseudo-critical points starting with $m_0$ almost zero and a circular border, a) $x_c^d$ and b) $x_c^o$.
Times are (from left to right) 1, 100, 1000, 10000 and 20000.
}
\end{figure}

It seems that the coarsening differs at the two pseudo-critical points, so as a final test the time 
evolution for a special initial condition were performed setting $m_0$
really close to zero putting almost half of the spins in one state and the other half in the other state  with a
circular border, in this way we have a large number of sites with $\beta_4$ while the number of possible
sites with $b_2$ increases (Fig.~\ref{snapshot_cir}). We can observe that the circular shape last longer at 
$x_c^o$.
In order to corroborate the effect of the difference between temperatures it should be performed simulations 
in spin-like systems for different ratios $\beta_4/\beta_2$, except for the equilibrium case $\beta_4/\beta_2=1$.

\section{Conclusion}
In this work it has been shown that the short time dynamic in the majority vote model
presents power law behavior at different control parameters for the growing, $x_c^d=0.84860(10)$,
and the decay processes, $x_c^o=0.85007(6)$. These pseudo-critical points are compatibles with results
for the critical point reported  previously, $x_c=0.850(2)$.  It has been show also that the dynamic 
in both cases is slower that in Ising model for all the quantities calculated ($m,~m^{(2)}$ and $A$).
These results seems to be related to the competing dynamic between the interface ($\beta_2$)
and bulk ($\beta_4$) temperatures associated to the dynamic, and as consequence to the absence of detailed balance
in the system. 
In order to corroborate these results additional simulations must be carry on in systems without
detailed balance.
The dynamical critical exponent ($z$) and the static critical exponent ($\beta/\nu$) has been 
evaluated independently using a growing process, in both cases the results were close to the Ising ones.
For the decay process the $z$ exponent was evaluated using results from static simulations founding that
the value is different from the obtained in the growing process.

\section{Acknowledgments}
I wish to thank G. P\'erez for his useful comments. This work was supported by Conacyt M\'exico through
Grant No. 61418/2007.

\end{document}